\documentclass[conference]{IEEEtran}
\IEEEoverridecommandlockouts
\usepackage{cite}
\usepackage{amsmath,amssymb,amsfonts}
\usepackage{algorithmic}
\usepackage{graphicx}
\usepackage{url}
\usepackage{multirow}
\usepackage{tabularx}
\usepackage{textcomp}
\usepackage{xcolor}
\def\BibTeX{{\rm B\kern-.05em{\sc i\kern-.025em b}\kern-.08em
    T\kern-.1667em\lower.7ex\hbox{E}\kern-.125emX}}
\begin{document}

\title{5G Integrated Communications, Navigation, and Surveillance: A Vision and Future Research Perspectives}

\author{\IEEEauthorblockN{Muhammad Asad Ullah, Vadim Kramar, \\Ville-Aleksi Kaariaho, Vasilii Semkin}
\IEEEauthorblockA{
\textit{VTT Technical Research Centre of Finland}\\ Espoo, Finland}
\and
\IEEEauthorblockN{Davi Brilhante, Hamada Alshaer, \\Charles Cleary}
\IEEEauthorblockA{
\textit{ART, Collins Aerospace}\\
Cork, Ireland}
\and
\IEEEauthorblockN{Giovanni Geraci}
\IEEEauthorblockA{\textit{Universitat Pompeu Fabra,} \\
\textit{Telefónica Scientific Research}\\
Barcelona, Spain}
\thanks{This work has been accepted in IEEE 25th Integrated Communications, Navigation and Surveillance Conference, April 8-10, 2025, Brussels, Belgium. Copyright has been transferred to IEEE.}}

\maketitle

\begin{abstract}
Communication, Navigation, and Surveillance (CNS) is the backbone of the Air Traffic Management~(ATM) and Unmanned Aircraft System~(UAS) Traffic Management~(UTM) systems, ensuring safe and efficient operations of modern and future aviation. Traditionally, the CNS is considered three independent systems: communications, navigation, and surveillance. The current CNS system is fragmented, with limited integration across its three domains. Integrated CNS~(ICNS) is a contemporary concept implying that those systems are provisioned through the same technology stack. ICNS is envisioned to improve service quality, spectrum efficiency, communication capacity, navigation predictability, and surveillance capabilities. The 5G technology stack offers higher throughput, lower latency, and massive connectivity compared to many existing communication technologies. This paper presents our 5G ICNS vision and network architecture and discusses how 5G technology can support integrated CNS services using terrestrial and non-terrestrial networks. We also discuss key 5G radio access technologies for delivering integrated CNS services at low altitudes for Innovative Air Mobility~(IAM) and Advanced Air Mobility~(AAM) operations. Finally, we present relevant challenges and potential research directions for further studies.
\end{abstract}


\section{Introduction}
In crowded airspace, where thousands of aircraft are flying, safe aviation operation would not be feasible without the support of Communication, Navigation, and Surveillance (CNS) systems. These three domains complement each other and form the foundation for safe, secure, predictable, and sustainable aviation operations~\cite{2}. Each of these CNS domains has a different nature and works independently using distinct radio access technologies (RATs), hardware, and frequency spectrum. For example, C–communication plays a crucial role in facilitating data exchange between aircraft-to-aircraft, aircraft-to-ground, and ground-to-ground~\cite{3}. Similarly, N-navigation is essential for flight routing, and S-surveillance helps to detect and track the position of an aircraft. All three services are equally important for aviation and essential for any stage of any flight, and therefore, are heavily integrated into the aircraft systems and engaged in the Air Traffic Management~(ATM) operations, making them the backbone of aviation.

In the current CNS ecosystem, the three domains are fragmented and poorly integrated. Each domain uses different aircraft onboard hardware transceivers, diverse ground infrastructure and frequency spectrum including high frequency~(HF), very high frequency (VHF), ultra-high frequency (UHF), L-band, Ka, Ku and X bands~\cite{Martin}. This increases development and operational costs as well as inefficient spectrum use.

However, recent developments in the aviation domain brought new considerations. In the near future, thousands of new generations of small, highly automated Unmanned Aircraft~(UA) are expected to operate at Very-Low Level~(VLL), mainly below 500 feet Above Ground Level (AGL)~\cite{ICAO2023}. In Unmanned Aircraft Systems~(UAS), communications enable the command and control (C2) link between a UA and a Remote Control Station (RCS). With the increasing complexity of such operations such as Beyond Visual Line of Sight~(BVLOS), and especially in specific geographical areas, e.g., urban and restricted environments, and close to the controlled by ATM areas, it might be required to use a dedicated range of digital services to ensure the safety of UAS operations. Such a complex combination of services is called UAS Traffic Management (UTM)~\cite{ICAO2023} or U-space in Europe~\cite{EuropeanCommission2021}. The information exchange in those systems is expected to rely on telecommunication networks. Thus, CNS infrastructure is considered to be an essential enabler for UTM/U-space systems~\cite{ICAO2023,SESAR3JU2023}.

At higher altitudes than VLL, Vertical Take-off and Landing (VTOL) Capable Aircraft (VCA)~\cite{EASA2024_VCA,EASA2024_UAS} operations are expected to transport cargo and passengers. Initially, those operations are expected to be with a pilot onboard and within areas with well-developed infrastructure, such as urban areas. Such operations are known as Urban Air Mobility (UAM)~\cite{ReicheColleen2018,Kramar2021} operations. Gradually, the operations will spread to rural areas, becoming regional, inter-regional and even international. With the increasing level of automation, we expect unmanned VCA operations in the future. The VCA operations combined with the UAS operations form novel concepts that are known as Advanced Air Mobility (AAM)~\cite{NASA25} and Innovative Air Mobility (IAM)~\cite{EuropeanCommission2022} concepts. With the growing number of any types of aircraft operating at different levels and in different airspace classes and the increasing complexity of their operations, it is essential to expect the convergence of ATM and UTM systems and the transformation of the C, N, and S systems into one integrated CNS~\cite{EuropeanCommission2022,SESARJointUndertaking2024}. 


 
 The current CNS system, originally designed for legacy manned aircraft (MA) and ATM, uses multiple fragmented hardware and provides limited coverage at lower altitudes. Due to safety and operational complexity reasons, UAS regulations restricts the maximum allowed take off weight (MTOW) for UA. This presents a significant challenge for small UA to carry multiple fragmented hardware transceivers. These hardware are energy hungry and needs onboard batteries to power CNS system. This makes the traditional fragmented CNS system challenging for small UAs.
 
The continuous growth in UAS, coupled with the limited CNS coverage and availability of radio spectrum resources, demands a single innovative framework for sustainable and cost-effective integrated CNS solutions for ATM and UTM~\cite{5GATM,icao2021utm}. To address these challenges, we advocate leveraging 5G terrestrial networks (TN) and non-terrestrial networks (NTN) for developing a resilient integrated CNS (ICNS). Benefiting from the higher bandwidth, standardized spectrum and compact user equipment (UE), we expect 5G ICNS will bring interdependency between the three separate domains to take full advantage of cross-domain synergies.  5G ICNS will introduce a non-fragmented framework that enhances spectrum efficiency, improves aviation safety for both high-level and low-level altitude operations as well as public safety on the ground, optimizes airspace capacity, reduces fuel (or battery) consumption and lowers carbon emissions. Thus, 5G ICNS will enhance safety and sustainability, meeting the current and future ATM and UTM systems needs with improved capacity, performance, and latency than the legacy CNS systems. The main contributions of this paper are summarized as follows:

\begin{itemize}
  \item First, we present an up-to-date review of the current and under-development aviation technologies for CNS services. This review includes information from regulatory organizations, recent state-of-the-art publications, and advancements within the industry. 
  \item Second, we present a future-forward 5G ICNS vision that can work independently or co-exist with current aviation technologies. Notably, the envisioned 5G ICNS system is not intended to fully replace current systems but rather to complement and enhance their capabilities.
  \item Third, we technically align CNS services to corresponding 3GPP 5G TN and NTN systems and RATs. 
  \item Fourth, we provide future research directions that will motivate further studies based on our 5G ICNS concept.
\end{itemize}

The rest of this paper is organized as follows. Section~\ref{sec:sec2} presents the technical background. In Section~\ref{sec:sec3},  we present an overview of legacy CNS, and Section~\ref{sec:5G} discusses 5G state-of-the-art for aviation. Section~\ref{sec:sec4} discusses our 5G ICNS vision, including network architecture.  Section~\ref{sec:sec5} discusses the open challenges and future research perspectives. Finally, Section~\ref{sec:sec6} concludes this paper with final remarks.

\section{Technical Background}
\label{sec:sec2}
In this paper, to lay the foundation when discussing future 5G integrated CNS systems, we consider the legacy ATM as well as UTM/U-space systems that are currently uprising. Table~\ref{table:tab1} lists the key aviation acronym and their definitions.

\begin{table}[t!]
    \centering
    \caption{The list of key Aviation Acronyms and Definitions.}
    \resizebox{\columnwidth}{!}{\begin{tabular}{|l|l|}
        \hline
        \textbf{Acronyms} & \textbf{Defination} \\ \hline
        A-PNT & Alternative Positioning, Navigation, and Timing \\
        A2A & Aircraft-to-Aircraft\\
        A2X &  Aircraft-to-Everything\\
        A2G &  Air-to-Ground\\
        ACAS & Airborne Collision Avoidance System \\
       ACARS &  Aircraft Communication Addressing and Reporting System \\
        ADS-B & Automatic Dependent Surveillance-Broadcast \\ 
        ADS-C & Automatic Dependent Surveillance-Contract  \\ 
        ADS-L & Automatic Dependent Surveillance-Light  \\ 
        AeroMACS & Aeronautical Mobile Airport Communication System \\
       ATN & Aeronautical Telecommunications Network \\
        
        ATC & Air Traffic Control \\ 
        ATM & Air Traffic Management \\
        C2 & Command and Control link\\
        CNS & Communication, Navigation and Surveillance \\
       CNPC & Command and Non-Payload Communication\\
        CPDLC & Controlled Pilot Data Link Communication \\ 
        DME & Distance Measuring Equipment \\ 
        GNSS & Global Navigation Satellite System \\ 
        ICNS & Integrated Communication, Navigation and Surveillance \\
        ILS & Instrument Landing System \\
        INS & Inertial Navigation System\\
        LDACS & L-band Digital Aeronautical Communications System \\ 
MLat & Multilateration \\ 
        PSR & Primary Surveillance Radar \\ 
        SSR & Secondary Surveillance Radar \\ 
        TCAS & Traffic Alert and Collision Avoidance System \\
        TDoA &Time Difference of Arrival\\
        UAS &Unmanned Aircraft System\\
        UATS &Universal Access Transceiver System\\
       UTM & UAS Traffic Management\\
        VDL & VHF Data Link \\ 
        VHF & Very High Frequency \\ 
        VOR & VHF Omni-directional Range \\ 
        \hline
    \end{tabular}}
    \label{table:tab1}
\end{table}

\begin{figure}[t!]
\centerline{\includegraphics[width=\columnwidth]{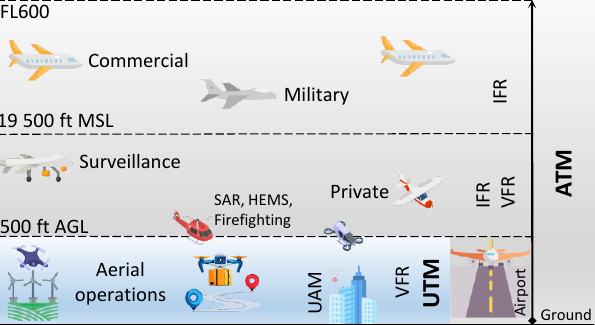}}
\caption{Illustration of aircraft operations supported by ATM and UTM.}
\label{fig:fig1B}
\end{figure}

Fig. \ref{fig:fig1B} illustrates a variety of aircraft operations supported with ATM and UTM systems. UTM is a separate framework to manage the UAS traffic at altitudes below 500 feet AGL. ATM remains responsible for the management of manned aviation at higher altitudes and controlled airspaces. Search and Rescue (SAR), firefighting, and Helicopter Emergency Medical Services (HEMS) operations should be supported by ATM but may penetrate the UTM operational areas. VCA may also operate with the support of ATM but penetrate the UTM areas, as shown in Fig. \ref{fig:fig1B}.  Equipped with advanced integrated CNS, ATM and UTM, are expected to benefit by obtaining data that can potentially enhance the safety of aviation\cite{Evgenii,icao2021utm}.  This requires specific means to allow UTM and ATM to exchange data. Those means are yet to be defined.

\subsection{Communication, Navigation, and Surveillance Service}
CNS services are crucial for a successful ATM and UTM operation. We define CNS following the definitions given in Regulation (EC) No 549/2004 of the European Parliament~\cite{EU_Regulation_549_2004}. 
\subsubsection{Communication service} It exchanges information, including the voice, messages and data for aeronautical fixed and mobile
services.

\subsubsection{Navigation service}  This provides an aircraft's precise location and timing information, which is essential for safe and efficient flight operations.  Navigation ensures an aircraft's safe and timely journey from departure to destination by providing accurate positioning and timing information throughout all the flight phases. 
\subsubsection{Surveillance service} It provides accurate information on aircraft position and mobility to allow safe separation between aircraft.

\begin{figure}[t!]
\centerline{\includegraphics[width=\columnwidth]{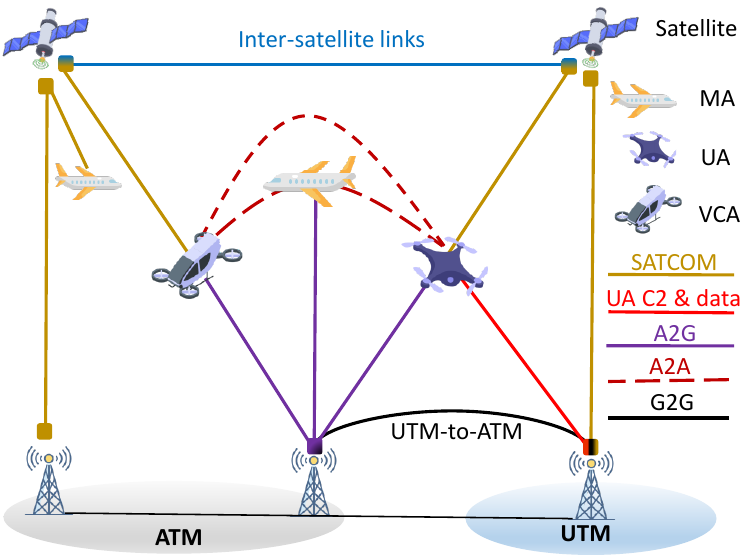}}
\caption{High-level illustrations of Aeronautical Telecommunication Network for providing CNS services to ATM and UTM.}
\label{fig:fig1}
\end{figure}

\subsection{Aeronautical Telecommunication Network}
Aeronautical Telecommunications Network (ATN) is a global aviation standard for aircraft-to-ground,  aircraft-to-aircraft, and ground-to-ground communication~\cite{ATN}. It is specifically designed to support air navigation service (ANS) operations by providing essential services. In Fig. \ref{fig:fig1}, we present a high-level illustration of an extended ATN system which can support CNS services for ATM and UTM and connects UTM-to-ATM using ground-to-ground communication data links.  Below, we discuss the key components of the ATN.

\subsubsection{Aircraft}  In the airborne segment of ATN, aircraft (including MA, UA and VCA) is equipped with wireless transceivers that transmit or receive data either to other aircraft or to the ground networks, which are further connected to ATM and UTM through aircraft-to-aircraft (A2A) or aircraft-to-ground (A2G) communication links, respectively. UA requires an essential dedicated C2 link for remote operation. 
\subsubsection{Access Network and Technologies}
The aircraft periodically broadcasts the messages that the nearby aircraft can receive for situational awareness. These messages include aircraft position, altitude and speed information. The aircraft-to-aircraft communication does not significantly rely on the terrestrial network infrastructure or satellite. However, a Ground-Based Augmentation System (GBAS) or Satellite Based Augmentation System (SBAS) could improve and ensure the aircraft's position accuracy. Ground-to-ground communication can be established either using a dedicated wired or wireless connection. Air-to-ground communication can either use TN or NTN, i.e., satellites, or even both. Typically, aircraft use satellite data links, known as SATCOM, in oceanic and remote areas outside the TN coverage. There are different data link options available, which can be used for delivering CNS services to ATM and UTM. Each has its pros and cons when it comes to UAS communication.
\begin{itemize}
\item \textbf{VDL:}  The band 117.975–137 MHz named VHF Digital Link (VDL) is one widely used for aviation communications which connect aircraft with ground network \cite{VDL_2024,VDL_2025}. Many legacy aviation protocols, including Controller Pilot Data Link Communications (CPDLC) and Aircraft Communication Addressing and Reporting System (ACARS), use VDL. The International Civil Aviation Organization (ICAO) defines four modes, named VDL Mode~1, VDL Mode~2, VDL Mode~3, and VDL Mode~4. VDL Mode~2 is the most widely used. It operates across 12 channels within the 136.700 – 136.975~MHz band with 25~kHz spacing. For aircraft-to-ground communications, both ACARS and CPDLC use VDL Mode~2. The UA and VCA can also use VDL links to communicate with the remote pilot, ATM and UTM \cite{Magali,icao2020uas}. However, small UA may have difficulty carrying the VDL module due to the small form factor. Due to the limited energy resources and equipment onboard, it is extremely challenging for small UA to continuously monitor the guard frequency bands 121.5~MHz to detect an emergency situation. Additionally, VHF frequency is shared among all aircraft within
range, in the near future, it will be nearly impossible for the limited VHF resources to accommodate the massive number of aircraft. Future ATM and UTM systems demand new connectivity solutions for CNS services. This motivated us to look at 5G for providing CNS services
\item \textbf{UAT:} The Universal Access Transceiver (UAT) operates at 978 MHz with a 1 MHz bandwidth, supporting the data link technology that enables Automatic Dependent Surveillance-Broadcast (ADS-B) functionality. UAT transceivers broadcast and receive important information, increasing situational awareness in the airspace \cite{UAT_2022}. Similar to manned aircraft, we expect that the UA and VCA can also use UAT to enable the surveillance capability \cite{icao2021utm}. However, UAT may pose similar shortcomings and challenges to UAS as VDL.
  \item  \textbf{Satellite:} In oceanic and remote environments, satellite communication is an important component of aeronautical communications. Satellites act as a bridge between aircraft and ground infrastructure to support aircraft-to-ground data links for ATM and UTM operations \cite{VDL_2024}. For instance, Viasat Classic Aero provides communication and surveillance services to aircraft for packet data including ACARS and ADS-B, and voice. With the rapid growth of the satellite communications industry, satellites will soon be able to cover urban areas, including crowded cities, while complementing TN during congestion and outages. In \cite{icao2021utm}, a combination of terrestrial and satellite-based communications is also recommended for UAS. However, this may require UAS to carry two separate hardware transceivers to use TN and NTN systems. Due to the small form factor, the UA may not be able to use traditional satellites, which require directional antennas.
  \item \textbf{ Long-Term Evolution:} 4G LTE is a 3GPP standardised technology. It has the potential to meet the needs of ATM and UTM systems. 4G offers better capacity and data rats than VDL and UAT systems. To give an example of a 4G-like system in aviation, L-band Digital Aeronautical Communications System (LDACS) uses Orthogonal Frequency-Division Multiplexing (OFDM) modulation to support digital data links between ground stations and aircraft for ATM functions \cite{LDACS_2018,LDACS_2021}. It operates on an L-band using 960 to 1164 MHz and offers throughput in the range of 550 kbps up to 2.6 Mbps, depending on the code rate and modulation scheme. Notably, LDACS has the potential to support navigation through Alternative Positioning, Navigation, and Timing (A-PNT). However, LDACS is mainly designed for the manned aircraft and ATM system.
\end{itemize}
Unlike these technologies, the Aeronautical Mobile Airport Communication System (AeroMACS) is built upon the Institute of Electrical and Electronics Engineers (IEEE) 802.16 standard, commonly referred to as Wireless Worldwide Interoperability for Microwave Access (WiMAX) \cite{AeroMACS}. AeroMACS is designed to support data links at the airport surfaces \cite{ATN}. It uses a channel bandwidth of 5 MHz in the 5091~MHz – 5150~MHz band and provides a throughput of 7.5~Mbps. Similar to LDACS, AeroMACS is mainly designed for ATM applications.
\subsubsection{Ground Infrastructure}  The ground infrastructure connects aircraft and the ground network, providing CNS services to ATM and UTM. It can also provide navigational aids, e.g., GBAS, to aircraft and receive surveillance information from them. Ground networks play a key role in ATM and UTM data exchange, as illustrated in Fig. \ref{fig:fig1}.

\if{0}
\textcolor{red}{
\subsection{Hyper Connected ATM}
Hyper-connected ATM is a new concept in aeronautical telecommunications networks, and it suggests the use of public and commercial communications systems (e.g., 5G) to support future ATM and U-space operations. Hyper-connected ATM will deliver high data rate broadband connectivity to provide additional support to ATM \cite{HyperConnected}. It will also offer internet connectivity to onboard passengers for entertainment during flight. In this paper, we discuss the 5G ICNS framework, which can be deployed using a shared (public) or a dedicated (private) spectrum.}
\fi
\section{An Overview of Legacy CNS Technologies}
\label{sec:sec3}
This section discusses the legacy CNS technologies of manned aircraft and their role during different phases of a flight operation. Additionally, it highlights the limitation of the legacy fragmented CNS system when it comes to use in UAS. We present legacy CNS technologies to gain a deeper understanding of how the CNS system works. This knowledge serves as a foundation for envisioning the 5G ICNS system in the later sections.
\subsection{Key Technologies and Protocols}
The legacy CNS system is fragmented and uses different technologies, hardware types of equipment and protocols that work independently to deliver communication, navigation, and surveillance services \cite{Martin,Dave}. 
\begin{table}[t]
    \centering
    \caption{Comparison of existing aviation protocols for communication service \cite{Martin,Dave, ATN}.}
    \resizebox{\columnwidth}{!}{
    \begin{tabular}{|l|c|c|c|}
        \hline
        \textbf{} & \textbf{Voice} & \textbf{CPDLC} & \textbf{ACARS} \\ 
        \hline
        \multirow{2}{*}{\textbf{Usage}}& Audio & Text & Flight data\\ 
        &  & message & Text message \\ 
        \hline
     \multirow{2}{*}{\textbf{Users}} & Cockpit, & Cockpit, & Aircraft, 
 the cockpit,\\
     & ATC & ATC & cabin crew, ATC \\
     
     \hline
      \multirow{3}{*}{\textbf{Contents}} & Clearance, & Clearance, & Weather updates, \\
      & requests, & request, &  equipment health,\\
       & other & other & flight plans\\
       \hline
        \textbf{Frequency Band} & HF, VHF & VHF & VHF \\\hline
        \textbf{Signal} & Analog & Digital & Digital \\\hline
        \textbf{Data Rate} &  -& 30 kbps & 2400 bps \\\hline
        \textbf{Periodic} &  &  & \checkmark \\ \hline
        \textbf{Unicast} & \checkmark & \checkmark  & \checkmark \\ \hline
         \textbf{Broadcast} & \checkmark &  &  \checkmark\\ \hline
    \end{tabular}
    }
    \label{table:tabCom}
\end{table}

\begin{table*}[htbp]
    \centering
    \caption{Comparison of existing aviation protocols and technologies for navigation service \cite{Martin,Dave,ATN}.}
    \resizebox{\textwidth}{!}{
   \begin{tabular}{|l|c|c|c|c|c|}
        \hline
        \textbf{} & \textbf{GNSS (GPS)} & \textbf{DME} & \textbf{VOR} & \textbf{ILS } & \textbf{NDB}\\ 
        \hline
        \multirow{2}{*}{\textbf{Usage}} & Positioning& Distance & Bearing & Approach & Bearing\\
        & Time &  & & guidance &  \\

        \hline
        \textbf{Sender} & Satellite & Ground Station & Ground Station  & Ground Station & Ground Station \\
        \hline
        \textbf{Receiver} & Aircraft & Aircraft& Aircraft  & Aircraft& Aircraft\\
        \hline
        \multirow{2}{*}{\textbf{Carrier frequency}} &  1575.42 MHz, 1227.6 MHz & 960-1215 MHz & 108.0-117.95 MHz  & 75, 108.10-111.95 MHz & 190-1750 kHz\\
        &  &  &   &  328.6-335.4 MHz &\\
        \hline
        \textbf{Signal} & Digital & Morse code & Morse code & Morse code & Morse code\\
        \hline
       \textbf{Techniques} & TDoA & Propagation delay & Phase shift & Signal strength & Angle of arrival\\

                \hline
        \textbf{Broadcast} & \checkmark &  & \checkmark & \checkmark & \checkmark \\ 
                \hline
        \textbf{Interrogation } &  & \checkmark &  &  &  \\ 
                \hline
    \end{tabular}%
    }
    \label{table:tabNav}
\end{table*}

\begin{table*}[htbp]
    \centering
    \caption{Comparison of existing aviation protocols and technologies for surveillance services \cite{Martin,Dave,ATN}.}
\resizebox{\textwidth}{!}{
    \begin{tabular}{|l|c|c|c|c|c|c|c|}

        \hline
        \textbf{} & \textbf{ACAS} & \textbf{ADS-B} &\textbf{ADS-C} & \textbf{ADS-L}  & \textbf{FLARM} &  \textbf{PSR} & \textbf{SSR}\\ 
        \hline
        \multirow{2}{*}{\textbf{Usage}} & Collision& Flight & Flight & Flight & Collision &  Detection & Detection\\
      & avoidance& information & information& information & avoidance& positioning & positioning\\
    \hline
     \multirow{1}{*}{\textbf{Sender}} & Aircraft& Aircraft& Aircraft & Aircraft& Aircraft& Ground station& Ground station\\
     \hline
        \multirow{2}{*}{\textbf{Receiver}} & Aircraft &  Aircraft& Ground station  & Aircraft& Aircraft& Ground station& Aircraft\\
        & Ground station& Ground station&   & Ground station& Ground station & & Ground station\\
        \hline
        \textbf{Frequency} & 129-137 MHz & 978, 1090 MHz & Variable & 868.2, 868.4 MHz & 868 MHz& -& -\\
        \hline
        \textbf{Signal} & Digital & Digital& Digital & Digital & Digital & Analog &Digital\\
        \hline
        \textbf{GNSS-based} & \checkmark & \checkmark & \checkmark & \checkmark & \checkmark &  &\\
        \hline
        \textbf{Broadcast} & & \checkmark &  & \checkmark & \checkmark & \checkmark &  \\
        \hline
        \textbf{Unicast} &  &  &  & \checkmark & &  & \\
       \hline 
       \textbf{Interogration} & \checkmark &  &  &  & &  & \checkmark \\
                \hline
    \end{tabular}%
    }
    \label{table:tabSur}
\end{table*}

\subsubsection{Communication} There are three commonly used technologies that connect cockpit (pilot)/aircraft with ATM by exchanging voice, text messages, and flight data \cite{ATN}. First, voice communication uses amplitude modulation and HF/VHF bands for clearance, requests and exchange of any other information. Second, CPDLC connects pilots with ATM and allows information exchange by text messages \cite{Martin}. CPDLC reduces the dependency on voice communication, which leads to readback/hear-back error reduction. Third, ACARS uses VHF for information exchange related to weather information, equipment on broad's health status, flight plans, and connecting flight status \cite{Dave}. Table~\ref{table:tabCom} provides the summary of these three technologies.

Similar to the pilot of a manned aircraft,  the remote pilot of a UA has the same ultimate responsibility for the safe operation of the aircraft and the safety of people on the ground \cite{icao2020uas}. UAS and unmanned VCA communications require reliable data links to connect remote pilot/RCS with UA and VCA using C2 links.

\subsubsection{Navigation}  In the CNS ecosystem, there are multiple technologies which either directly support aircraft navigation in airspace or complement it by providing navigational aids (navaids). Today, the Global Navigation Satellite System~(GNSS) is the most common and widely adopted system for both manned and unmanned aircraft. The manned aircraft also use beacon-based navigation solutions, such as distance measuring equipment (DME), very high-frequency omnidirectional range (VOR), instrument landing system (ILS), and non-directional beacon (NDB)  are either used separately or in a combination to complement GNSS-based navigation or provide A-PNT support when GNSS signals are unavailable \cite{Martin,Dave,ATN}. Table~\ref{table:tabNav} summarises the legacy navigational technologies.

Most of the manned aircraft navigation techniques, including DME, VOR, ILS, and NDB, may not work for small UAS due to their form factor, limited energy resources, and MTOW limitations. Therefore, there is a vital need for a robust UAS navigation framework that can complement GNSS and work as an alternative in case of GNSS outage, jamming or spoofing. GBAS and SBAS systems are often used to increase the accuracy of GNSS positioning. However, these traditional systems and hardware may need modifications when it comes to UAS augmentation. As an alternative, post-processed kinematic (PPK) and real-time kinematic positioning (RTK) are commonly used to correct GNSS systems and provide centimeter-level accuracy in UAS positioning~\cite{RTK}. In the navigation context, geofencing is another technology to avoid no-fly-zone~(NFZ)~\cite{Geofence}. Similarly, inertial navigation systems~(INS) support UAS navigation and facilitate maintaining accurate flight paths.

\subsubsection{Surveillance} Finally, this third service provides accurate information on the position and movement of aircraft to ATM and UTM systems for ensuring safe separation between aircraft. In manned aviation, the commonly used surveillance technologies include Airborne Collision Avoidance System~(ACAS), ADS-B, Automatic Dependent Surveillance - Contract (ADS-C), Primary Surveillance Radar (PSR), and Secondary Surveillance Radar (SSR). Together, these technologies improve situational awareness and contribute to the effective management of manned airspace.  Table~\ref{table:tabSur} summarises the typical surveillance technologies.

To avoid UAS mid-air collisions, UA must feature Detect and Avoid (DAA) technology onboad.  ACAS is a DAA system, and it works independently of the ATM system. The legacy ACAS system is designed for manned aircraft, which must be able to achieve a rate of climb of 2500 ft/min \cite{eurocontrol2022acas}. Therefore, small UA do not have an ACAS system  \cite{icao2020uas}.  ACAS Xu is the latest version of ACAS designed for UAS; it enables UA with DAA capabilities \cite{ACASXU}. The UAS Remote ID broadcasts UA identification and location information for surveillance\cite{RemoteID}. 

The UA can be equipped with ADS-B and ADS-C,  Automatic Dependent Surveillance-Light (ADS-L) and FLARM to provide situational awareness and surveillance services to UTM. Radars, e.g., SSR and wide area multilateration (WAM), can also be used for the UAS surveillance \cite{icao2021utm}.  PSR might be less accurate in tracking smaller UA at low altitudes.

\subsection{Usage during flight phases}
\begin{figure}[t!]
\centerline{\includegraphics[width=\columnwidth]{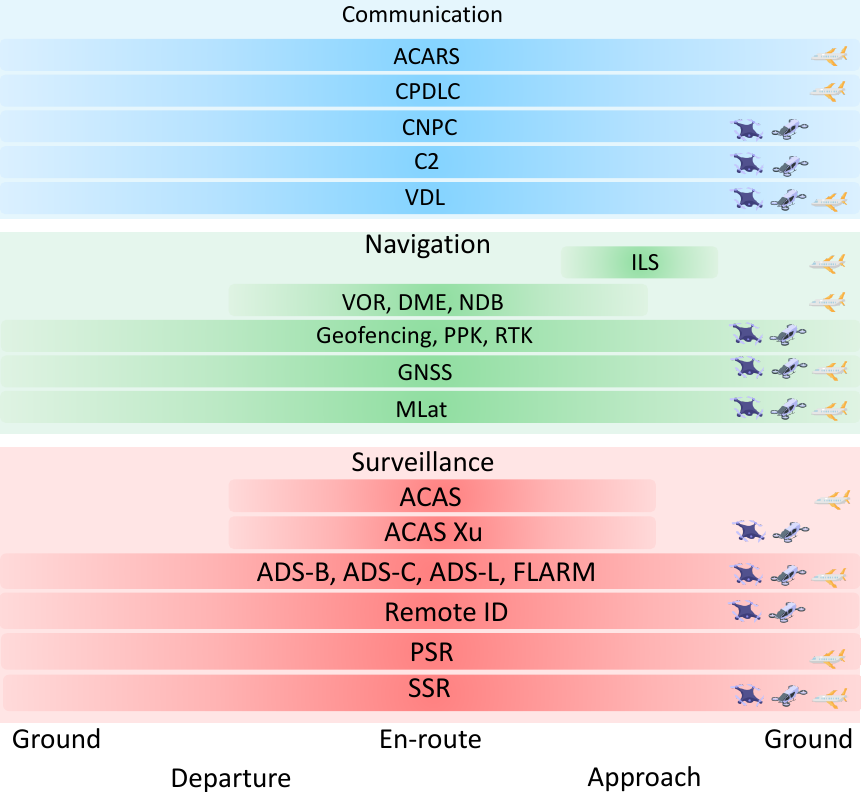}}
\caption{CNS technologies usage during the different flight phases of manned aircraft and UAS.}
\label{fig:fig3}
\end{figure}

The use of CNS services varies across different phases of a flight, as shown in Fig.~\ref{fig:fig3}. Manned aircraft use ACARS, CPDLC, and VHF-based VDL communication systems throughout all flight phases. UAS uses C2, CNPC, and VHF-based VDL communication during all phases, where VDL is the least suitable option.

Navigation technologies GNSS and Multilateration (MLat) provide precise positioning and timing from departure to arrival and can be used for both manned aircraft and UAS. Manned aircraft use DME, NDB, and VOR beacons that assist with navigation during the en-route phase, while the ILS system guides the aircraft during its approach to the runway~\cite{Martin}. UAS needs geofencing, PPK, and RTK for navigation during all flight phases.

Similarly, surveillance systems, i.e., ADS-B/ADS-C/ADS-L, FLARM, PSR, and SSR, are required during every stage of a flight. ACAS and ACAS Xu (designed for UAS) play the most critical between departure and approach, called the en route phase. Typically, Remote ID broadcasting is required throughout the entire UAS flight. 

\subsection{CNS for Very-Low Level operation}

\begin{table}[t]
    \centering
    \caption{CNS technologies and their applicability to manned and unmanned aircraft systems.}
    \resizebox{\columnwidth}{!}{
    \begin{tabular}{|l|c|c|c|}
        \hline
        \textbf{Technology} & \textbf{Manned} & \textbf{Unmanned} & \textbf{Category} \\
        & \textbf{Aircraft} & \textbf{Aircraft} & \textbf{} \\
        \hline
      \textbf{ACARS} & \checkmark  &   & Communication\\ \hline
       \textbf{VDL} & \checkmark  & \checkmark  & Communication\\ \hline
       \textbf{C2} &   &  \checkmark & Communication\\ \hline
       \textbf{CNPC} &   &  \checkmark & Communication\\ \hline 
      \textbf{CPDLC} & \checkmark  &   & Communication\\ \hline
     \textbf{Geofencing} &  & \checkmark & Navigation   \\ \hline
    \textbf{INS} & \checkmark  &   \checkmark & Navigation\\ \hline
       \textbf{GNSS} & \checkmark & \checkmark &  Navigation, Surveillance\\ \hline
  \textbf{PPK}  &  & \checkmark &  Navigation, Surveillance\\ \hline
\textbf{RTK} &  & \checkmark &  Navigation, Surveillance\\ \hline
      \textbf{A2X} &   & \checkmark  & CNS\\ \hline
      \textbf{ACAS} & \checkmark  &   & Surveillance\\ \hline
     \textbf{ACAS Xu} &  & \checkmark & Surveillance\\ \hline
        \textbf{ADS-B} & \checkmark  & \checkmark  & Surveillance\\ \hline
     \textbf{ADS-C} & \checkmark  & \checkmark  & Surveillance\\ \hline
      \textbf{ADS-L} & \checkmark  & \checkmark  & Surveillance\\ \hline
    \textbf{DME} & \checkmark &   & Surveillance \\ \hline
    \textbf{ILS} & \checkmark &   & Surveillance \\ \hline
    \textbf{NDB} & \checkmark &   & Surveillance \\ \hline
    \textbf{VOR} & \checkmark &   & Surveillance \\ \hline
        \textbf{PSR} & \checkmark &   & Surveillance \\ \hline
       \textbf{Remote ID} & &  \checkmark & Surveillance \\ \hline
        \textbf{SSR} & \checkmark &  \checkmark & Surveillance \\ \hline
    \textbf{TCAS} & \checkmark &   & Surveillance \\ \hline

    \end{tabular}%
    }
    \label{table:UAS_CNS}
\end{table}

\begin{table*}[t!]
    \caption{Summary of 3GPP state-of-the-art for UAS.}
    \resizebox{\textwidth}{!}{
\begin{tabular}{|l|c|c|c|c|c|}
    \hline
\textbf{Doc.}         & \textbf{Rel.} & \textbf{Description} & \textbf{C} & \textbf{N} & \textbf{S} \\
\hline
\multirow{2}{*}{TR 38.777}    & \multirow{2}{*}{15}   & Establishes the radio level KPIs and basic UAS operation features, evaluating LTE ground users coexistence & \multirow{2}{*}{\checkmark }  &   &   \\
&    &  with UAS users &   &   &   \\
\hline
TR 22.825    & 16   & Initial architectures and networking services definitions &  \checkmark  & \checkmark   &   \\
\hline
TS 22.125    & 16   & UAS system requirements for data applications  & \checkmark   & \checkmark   &   \\
\hline
TR 22.829    & 16   & 5G networked UAS use-cases & \checkmark  &   &   \\\hline
TR 23.754    & 17   & Identification and tracking procedures for UAS services & \checkmark   &   &   \\\hline
TR 23.755    & 17   & Application architecture description to support efficient UAS operations & \checkmark   &  \checkmark  &   \\\hline
\multirow{2}{*}{TS 23.256}    & 17   &  Architecture enhancements for supporting the communication needs of UAS, identification, and tracking, & \checkmark    & \checkmark   &\checkmark   \\
&   &  according to the use cases and service requirements & \checkmark    &    &  \\\hline
TS 23.255    & 17   &  Functional architecture, procedures and information flows for UAS application & \checkmark   &  \checkmark  &  \checkmark  \\ 
\hline
TS 24.578 & 18 & A2X services access to 3GPP network functions through the PC5 interface & \checkmark   & \checkmark   & \checkmark  \\\hline
\end{tabular}
}
 \label{table:SotA}
\end{table*}

The combinations of continuous coverage, high bandwidth, and precise timing and position are crucially needed for UAS operation at VLL.   Existing CNS infrastructures and technologies are primarily designed for commercial and general aviation to offer services in designated areas, including airports and air corridors.  Table \ref{table:UAS_CNS} shows the CNS technologies and their adaptability to manned aircraft and UAS.

 The majority of current CNS technologies need perfect line-of-sight (LOS). However, the probability of LOS between aircraft and ground stations decreases when operating at VLL. In the case of urban operations, the building can also potentially block the LOS and restrict the operation.  For example, the signal quality of GNSS satellites is poor in urban environments due to the obstruction of buildings, which prevents navigation services. However, PSR may not be able to track small UA at VLL accurately. Given this, CNS coverage remains limited and challenging at VLL~\cite{VLL}. There remains a need for new solutions supporting CNS services for manned aircraft and UAS at VLL.
\section{5G and UAS Integration: State-of-the-Art}
\label{sec:5G}
In September 2016, the Radio Technical Commission for Aeronautics (RTCA) published the Minimum Operational Performance Standards (MOPS) for UAS  Command and Non-Payload Communication (CNPC) \cite{VLL,NASA_CNPC}. According to the ITU-R report (ITU-R M.2229 \cite{ITU2011}) CNPC bandwidth requirements are 34 MHz and 56 MHz for the terrestrial-based LOS CNPC and for the satellite-based beyond LOS (BLOS) CNPC link, respectively \cite{Hosseini,Andrew}. 

3GPP 5G TN and NTN are promising technologies to meet these requirements \cite{Geraci}. Table \ref{table:SotA} summarises the 3GPP state-of-the-art for UAS. The 3GPP support to UAS dates to 2018, when the first studies were drawn \cite{tr38.777} in Release (Rel.) 15. At first, the objectives were to identify the capability of serving such aircraft, evaluate the performance, and enhance solutions to optimize LTE connectivity for UA. Mainly, at this stage the radio level KPIs and features were being defined to support the UAS applications. The three CNS domains were not defined as individual requirements, but they fit application data categories, which are specified in conjunction with C2 for the UA piloting and real-time safety systems. Then, C2 and data will demand different reliability (10$^ {-3}$ for C2), latency (50 ms one way for C2) and data rate (100 kbps for C2 and 50 Mbps for application data) requirements. This study concludes that LTE support for a small number of aircraft and ground users is acceptable but was challenging in terms of mobility and interference control for areas with a higher density, requiring specification enhancements.

With the evolution to 5G NR and its improved network capabilities, new requirements were specified in \cite{ts22.125} for UAS and UTM, including data application, C2, and positioning requirements. Several UAS use-cases and their requirements are defined in \cite{tr22.829}, such as UA high-resolution video live broadcasting, C2 communication, UA and ground eMBB users’ coexistence, autonomous UAS controlled by artificial intelligence (AI), UAS in logistics, and UAS controller change. 3GPP Technical Report (TR) 22.829 also contains the necessary KPIs, such as use cases and general UAS-bsed support services, like telemetry, laser mapping, and video streaming.

More concrete UAS architectures and networking services are provided in TR~22.825~\cite{tr22.825}, which contains infrastructure, scenario description, and requirements for multiple UAS issues, such as initial authorisation, data acquisition, identity broadcast, NLOS operation, and cloud operation. In TR~22.825, the basic set of information that describes a UAS and its main characteristics, like UE capabilities and device identity, are defined. In Rel. 17, the 3GPP starts focusing on the network architectures and functionalities to support UAS operation. Some navigation and surveillance functions have been mentioned, including aircraft geofencing, DAA, and UAS flight path-restricted area exposure. Specifically, the TR~23.754~\cite{tr23.754} sheds light on the UAS and its controller identification and tracking procedures. It specifies the network infrastructure and processes to support different UAS configurations, such as UA-to-UA communication and network-assisted C2 communication. Complementary to this, TR~23.755~\cite{tr23.755} specifies the application architecture to support efficient UAS operations over 5G networks. Several issues have been raised regarding the UAS network application, and architectural solutions have been proposed to mitigate such issues, such as positioning support, C2 QoS provisioning, switching and selecting C2 communication modes, and real-time UA connection monitoring and location reporting.
\begin{figure*}[t!]
\centering
\includegraphics[width=\textwidth]{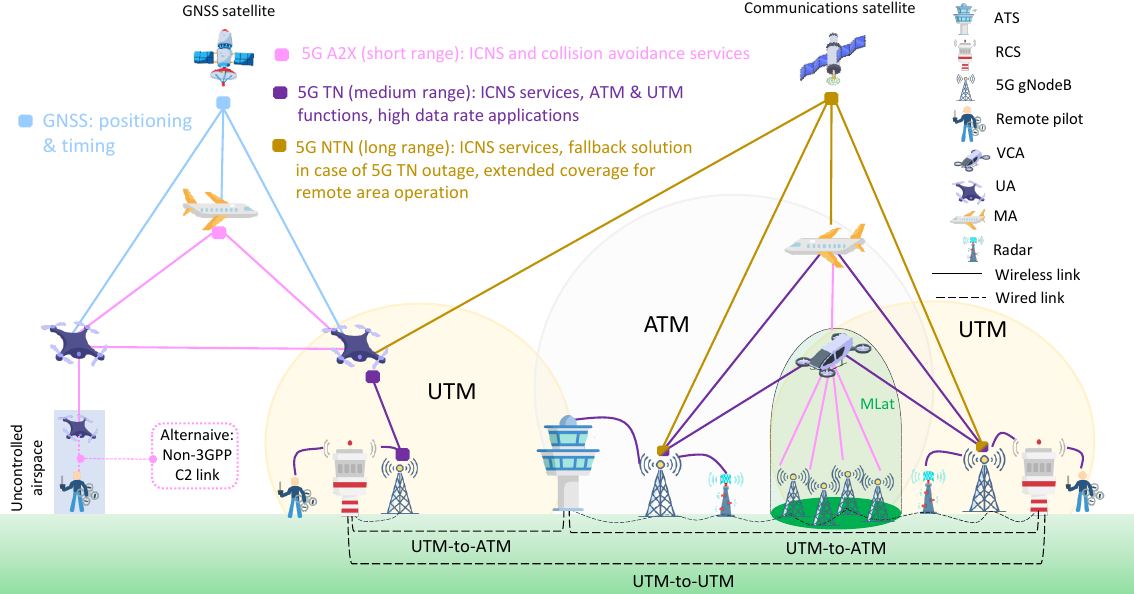}
\caption{Illustration of our 5G integrated CNS vision.}
\label{fig:fig4}
\end{figure*}
The most recent standardisation efforts in Rel.18 define the main interfaces to allow direct access to the 3GPP network functions through the PC5 interface, resulting in better integration between UAS services and 5G networks, also denominated as A2X (aircraft to everything) services. In \cite{ts24.578}, the key integration issues are raised, aiming to identify the architectural and functional modifications required to support UAS. Such study includes enabling C2 direct communication in a UAS and detecting and avoiding technology for collision prevention using the Uu and PC5 5G NR standard interfaces. The PC5 interface was originally designed to allow communication between vehicles and roadside units but is being extended to enable UAM. This interface provides, for instance, collision avoidance through traffic-safety messages \cite{namuduri2022advanced }. The TS~23.256~\cite{ts23.256} and TS~23.255~\cite{ts23.255} provide a description of such interfaces and the overall capabilities of UA as a UE, such as registration, handover, and Packet Data Unit (PDU) session establishment.

In academia, the perception is that CNS technologies must improve to meet future aviation demands, particularly low-altitude aircraft operations. Due to its size, the small UA can not embed the plethora of instruments that exist in larger aircraft operating at higher altitudes. To address this challenge, 5G and future generations are envisioned to fill this gap using TN and NTN systems \cite{erturk2020requirements}. The integration between 5G TN and NTN also tackles the low coverage challenge in remote and isolated areas for UAS operations that are far from urban scenarios. Finally, adding essential functionalities for flight, such as DAA and identity broadcasting, in the 3GPP 5G NR releases consolidates it as an emerging and evolving technology for CNS support.

This paper advocates for leveraging 5G technology to provide CNS services to any aircraft operations to ensure service availability and reliability at any altitude of the operation within the VLL and low level.

\section{Our Vision: 5G Integrated CNS}
\label{sec:sec4}
The network infrastructure and radio spectrum for Communications (C), Navigation (N), and Surveillance (S) play a critical role in transmitting the traffic generated by these services and directly influence overall performance. In a fragmented (non-integrated) CNS network infrastructure, each service generates its separate stream of data packets, creating separate traffic flows, and often uses a separate network and dedicated hardware for the transmission. This fragmented CNS network architecture results in different quality-of-service~(QoS) for each service. Additionally, it is worth reminding it is challenging for small UAS to carry multiple fragmented transceivers and sufficient battery to accommodate several CNS systems.

In this paper, we present our ICNS concept which uses a single integrated framework with minimal onboard hardware. Specifically, we advocate leveraging 5G TN-NTN technologies and a single multimode transceiver. Fig. \ref{fig:fig4}\footnote{Icons are made by multiple authors (Freepik, Smashicons, Prosymbols Premium, mynamepong, Konkapp, vectorsmarket15, 
Luvdat, Kanyanee Watanajitkasem, Triangle Squad, and Ylivdesign) from www.flaticon.com.} illustrates our 5G~ICNS vision.

\subsection{Integrated CNS}
By integrating these three domains, ICNS is expected to enhance the efficiency and resiliency of ATM and UTM operations, promote more efficient use of the scarce radio spectrum, and improve QoS for all airspace users. An integrated CNS framework will eliminate the need for multiple onboard hardware equipment, lowering the UA's payload. ICNS system implies the convergency of the previously separate CNS services. This requires a single integrated framework and a resilient network architecture with minimum hardware to merge the traffic flows of these three services and handle their packets according to the needs of required performance in a cost-efficient manner.

\subsection{5G ICNS Network Architecture}

Integrated TN and NTN provide resilient, continuous, and ubiquitous wireless coverage \cite{Geraci}. We propose leveraging 3GPP TN and NTN radio access technologies for delivering ICNS services.  The proposed 5G ICNS architecture will deliver communication, navigation, and surveillance services using the 3D network, supporting aircraft operations across large coverage areas and at all altitudes ranging from VLL to high altitudes. 
 
The role of the TN system will be to provide the primary connection in developed areas with ground network infrastructure. At the same time, NTN will support CNS operations in regions outside of TN coverage. Additionally, in urban areas, NTN can help to offload the CNS traffic from TN, reducing congestion during peak hours. In case of a TN outage, NTN will act as a fallback solution to ensure continuous connection availability and delivery of CNS services, strengthening resilience. For autonomous operations beyond visual line-of-sight, the envisioned 5G ICNS network architecture with integrated TN-NTN will increase connection availability, offer more considerable network coverage, ensure service 
continuity and improve resilience. We discuss the role and characteristics of different 5G links for ICNS services below.
\begin{figure*}[t!]
\centerline{\includegraphics[width=\textwidth]{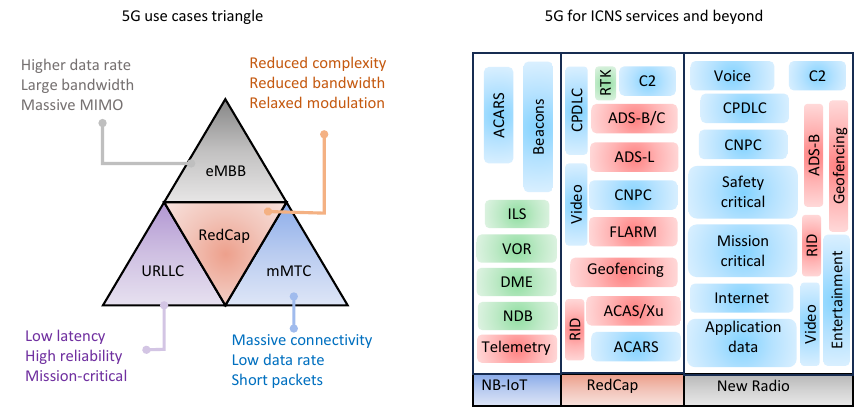}}
\caption{5G use cases triangle and envisioned radio access technologies for the ICNS services and beyond.}
\label{fig:fig5}
\end{figure*}
\subsubsection{5G Terrestrial Network} In the presence of ground networks covering, the 5G TN system will ensure essential CNS services. Similarly, it will also support safety and mission-critical communication and the internet connection for the passengers of VCS. 
\subsubsection{5G Non-Terrestrial Network} In the case of TN outage or congestion, 5G NTN will act as a fallback solution to ensure the ICNS system's resilience \cite{BenGerNik2022}. Similar to the TN system, 5G NTN can also provide the required connectivity. 5G NTN offers sufficient data rates that can support safety, mission-critical communication, and broadband connectivity for onboard passengers and flight crew\cite{Damini}. The nature of the LEO satellite payloads, whether regenerative or transparent and the presence of inter-satellite links may also influence the network configuration presented in Fig. \ref{fig:fig4}.
\subsubsection{5G Aircraft-to-Everything} A2X is a 5G sidelink technology introduced in 3GPP Rel. 18 \cite{3GPP_R18_A2X,ts24.578}. It is designed to enhance collision avoidance mechanisms, ensuring the safety and efficiency of aviation flights. In our proposed 5G ICNS system, we recommend the use of 5G A2X technologies for short-range CNS and collision avoidance service to enable UA situational awareness in the surrounding area. For example, A2X can support aircraft-to-aircraft and air-to-ground (short-range) communication for exchanging ADS-B, ADS-L, C2 and Remote ID, ACAS/ACAS-Xu messages. Additionally, A2X can also be used to broadcast the beacons, which could be further proceeded by the ground station for MLat purposes to ensure A-PNT, as illustrated in Fig. \ref{fig:fig4}. Such links can be underlaid with the conventional cellular uplink \cite{AzaGerGar2020} for spectral efficiency. However, further investigations are needed to assess the feasibility of this idea. 

In the envisioned ICNS framework, the 5G network can use TN, NTN or A2X to provide highly accurate timing signals to UAS. These timing signals can support RTK-based navigation and geofencing. Notably, when considering the development of wireless transceivers that could be incorporated into both manned aircraft and UAS, special attention is needed for the end device (5G user equipment–UE) capabilities and form factor. Notably, the size and weight of the payload should be appropriate. Such UE may support multiple radio access technologies and could work on different frequencies to benefit from the full potential of 5G TN-NTN for ICNS services.
\subsection{Multi-mode Radio Access
Technologies}

5G use cases include Ultra-Reliable and Low Latency Communications (URLLC), Enhanced Mobile Broadband (eMBB) and Massive Machine-Type Communications (mMTC) for the TN systems \cite{Petar,Asad,Asad2}. Due to high propagation delays and long link distances, URLLC is not feasible with NTN, especially satellites. IMT-2020 ITU-R M.2514-0 defines Highly Reliable Communication (HRC) as an alternative to URLLC for 5G NTN \cite{Mikko}. We envision an aircraft equipped with 5G multi-mode transceivers, allowing it to select the most suitable RAT based on the requirements of CNS services. Fig. \ref{fig:fig5} presents the 5G use cases eMBB, eMTC, and URLLC triangle and suitable 5G RATs for delivering ICNS services.
\subsubsection{5G New Radio} 5G NR is the main RAT that connects the payload/UE with the 5G base station (gNodeB) using the protocols specified by 3GPP. 5G NR supports eMBB and URLLC services and use cases. In ideal channel conditions, 5G NR peak data rates are 20~Gbps and 10~Gbps for downlink and uplink, respectively \cite{Nurul}. This makes 5G NR a prominent choice for aircraft safety/mission-critical communication and non-safety critical commercial communication, including internet connectivity for onboard passengers. 5G NR can provide high data rate communication to facilitate voice, video, internet and onboard entertainment services. In the context of UAS, 5G~NR can provide high data rate links for C2 between the RCS and UA and accommodate pilot voice communications. 5G NR can also meet the required data rates of multiple UAS applications defined in 3GPP TS22.125 \cite{ts22.125}. This includes 8K video live broadcast, laser mapping, HD patrol, 4K AI surveillance, real-time video, video streaming and image transmissions.
\subsubsection{5G RedCap}
The Reduced Capability (RedCap) is a new 5G RAT introduced in 3GPP Rel.~17 and further enhanced in Rel.~18 \cite{Champaka,DtS}.  RedCap UE can co-exist with the legacy 5G NR in the same network and can be served by the same gNodeB without hardware modifications. RedCap and enhanced RedCap support bandwidths of 20~MHz and 5~MHz, respectively, with peak rates of 220~Mbps and 10~Mbps, correspondingly. This throughput is at least ten times higher than that of the UAT and Mode-S Extended Squitter Transponder~(1090ES) systems, which are commonly used for ADS-B and ACAS/SSR, respectively. Similarly, the RedCap throughput is approximately 300 times more than the VDL~Mode~2 system which is used for CPDLC and ACARS transmissions. Due to better throughput and low latency, we recommend 5G RedCap/eRedCap as a potential RAT for delivering CNS services, as shown in Fig.~\ref{fig:fig5}. Similar to 5G NR, the Rel. 17 RedCap with 20 MHz bandwidth can support most of the UAS applications defined in 3GPP TS 122 125 \cite{ts22.125}. This also includes high-quality video streaming. Due to 5 MHz bandwidth, Rel. 18 eRedCap does not provide sufficient data rates to support demanding applications such as 8K live video broadcasting, laser mapping, high-definition patrol operations, and 4K AI-driven surveillance. eRedCap may support real-time video or image transfers in addition to C2.

\subsubsection{NB-IoT} Compared to 5G NR and RedCap, NB-IoT offers larger coverage up to 35~km (on-ground) in perfect LOS conditions and supports more users.  There is a strong industrial and academic interest in using NB-IoT for NTN~\cite{NB-IoT_NTN,Alessandro}. NB-IoT targets mMTC use cases that require low data rates. Specifically, it offers a maximum downlink data rate is up to 226.7~kbps and 250~kbps for the uplink.  Due to low data rates, NB-IoT cannot support UAS applications and C2 traffic. However, we consider NB-IoT as a suitable RAT for short data transmissions e.g., ACARS, telemetry (IoT-like short data indicating the health status of equipment on UA) and beacons for ISL, VOR, DME, and NDB, as shown in Fig.~\ref{fig:fig5}.  

\subsection{Frequency Spectrum}
\begin{figure}[t!]
\centerline{\includegraphics[width=\columnwidth]{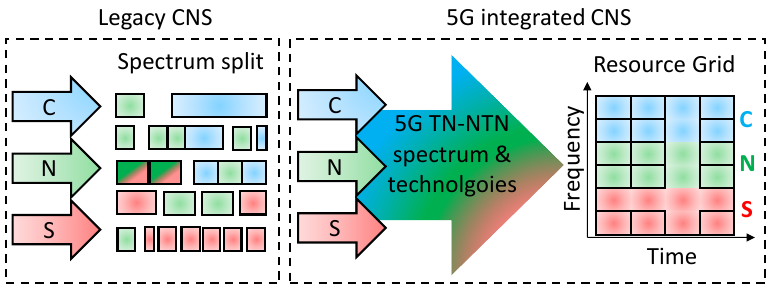}}
\caption{A high-level illustration of the frequency spectrum used by legacy CNS and envisioned integrated 5G ICNS framework.}
\label{fig:fig6}
\end{figure}

The design, development, and implementation of 5G ICNS network architecture, illustrated in Fig. \ref{fig:fig4}, requires significant synergies between TN and NTN systems. Our envisioned ICNS will use the radio access technologies, radio access standards, and radio frequency spectrum defined by 3GPP specifications for the 5G technology. Unlike legacy CNS, which uses multiple frequencies in HF, VHF, UHF, L-band, Ka, Ku, and X bands~\cite{Martin}, our envisioned 5G ICNS framework relies on 3GPP specified uniform spectrum. Fig. \ref{fig:fig6} illustrates how legacy CNS and our proposed integrated 5G ICNS framework use frequency spectrum.

A larger channel bandwidth is essential to achieve high data rates. A higher-frequency signal has a shorter wavelength, which generally means it is more susceptible to attenuation and may not travel as far as a lower-frequency signal with a longer wavelength. While communication coverage reduces as the frequency increases, path loss can be compensated via beamforming with larger antenna arrays. 
To ensure a balance between sufficient data rates and coverage, the proposed 5G ICNS system requires further research to mature the solution and strategically select suitable frequency bands and a network (TN, NTN, or A2X) for broader coverage and increased capacity. Below, the possible carrier frequencies and channel bandwidths are discussed.

\subsubsection{Terrestrial Network} There are three radio spectrum options available for 5G NR which include Frequency Range~1 (FR1, sub-7~GHz frequency bands), Frequency Range~2 (FR2 from 24.25~GHz to 71.0~GHz), and Frequency Range~3 (FR3, 7.125~GHz to 24.25~GHz). FR3, also known as upper mid-band, is envisioned as the new spectrum range for 6G~\cite{LopPioGer2025,FR3}. Terrestrial gNodeB often uses lower frequencies (800~MHz, 900~MHz, 1800~MHz, or 3600~MHz) for wider coverage.
\subsubsection{Non-Terrestrial Network}  NTNs typically use satellites in LEO,  Medium Earth Orbit (MEO), and Geosynchronous Orbit (GEO) to provide connectivity. Path and attenuation losses increase with frequency; therefore, the satellite industry prefers lower frequencies (L and S-band) for low data rate applications. 3GPP defines several NTN  frequency bands in FR1 and FR2 with channel bandwidths 5- 20~MHz and 50-400~MHz, respectively \cite{Champaka}.
\subsubsection{Aircraft-to-Everything} Federal Aviation Administration~(FAA) has approved the 5030-5091 MHz band to support UAS control links \cite{FCC2024}.   Qualcomm is requesting the FCC and advocating the same band for A2X communication with a channel bandwidth of 20 MHz \cite{Qualcomm2024}.

\subsection{Data 
Packets and Traffic Flow}

We suggest using IP-based data packets, routing algorithms, and network slicing to create logical data pathways specific for each CNS service. In the envisioned 5G ICNS concept, data packets are routed according to their QoS requirements, reflected in their required performance, through the ICNS network architecture to deliver the CNS data packets to their destinations. The CNS data can be transmitted either through three separate packets or within a single aggregated packet, using the proposed network architecture.

\subsection{5G-based A-PNT}
A CNS must be resilient against jamming and spoofing threats.  It is worth emphasising that both GNSS and ADS-B (which further uses position information derived from GNSS) are vulnerable to jamming and spoofing threats. Additionally, the GNSS signals are weak due to the significant path loss and not ubiquitous because of the satellite constellation designs. Similarly, the legacy ADS-B coverage is either poor or absent in oceanic/remote areas. As an alternative, satellite-based ADS-B solutions are available for the oceanic regions. To give one example, Aireon supports ADS-B in oceanic regions, which may not be able to handle the needs of growing aircraft in the future due to bandwidth limitations. Notably, Aireon uses the 1090ES onboard hardware and primarily targets the aircraft at high altitudes~\cite{ICAO2019}. 

To mitigate these challenges, our envisioned 5G ICNS system will use MLat-based A-PNT to ensure the accuracy, availability, continuity, and integrity of the services at VLL. Compared to GNSS signals, 5G signals are significantly stronger and more robust to jamming and spoofing threats. In case of GNSS degradation or outage, the envisioned 5G ICNS can support navigation by MLat. However, it will require at least four ground stations (gNodeB) to perform MLat, as presented in Fig. \ref{fig:fig4}.   

\section{Challenges and Future Works}
\label{sec:sec5}
\begin{table}[t]
\centering
\caption{Comparison of LDACS, VDL Mode 2, and envisioned 5G ICNS.}
\resizebox{\columnwidth}{!}{
\begin{tabular}{|c|c|c|c|}
\hline
\textbf{} & \textbf{LDACS} & \textbf{VDL Mode 2 } & \textbf{5G ICNS} \\ \hline
\multirow{1}{*}{Applications} & CPDLC, ADS-C& CPDLC,ADS-C & ICNS  \\ \hline
\multirow{2}{*}{Channels} & Very &Medium & Extremely \\
                            & high &to low & high \\ \hline
\multirow{2}{*}{Data rate} & Very & \multirow{2}{*}{Low} & Extremely\\ 
                            & high & & high\\ \hline
\multirow{1}{*}{Long-term} & Very & Very & Extremely \\ 
\multirow{1}{*}{capacity }                                  & high &  low &  high \\ \hline
\multirow{1}{*}{Navigation} & \multirow{2}{*}{\checkmark} &  & \multirow{2}{*}{\checkmark} \\
\multirow{1}{*}{capability} &  &  & \\ \hline
Secure & \checkmark &  & \checkmark\\\hline
\multirow{1}{*}{Voice} & \multirow{2}{*}{\checkmark} &  & \multirow{2}{*}{\checkmark}\\
\multirow{1}{*}{capability } &  &  & \\ \hline
\multirow{1}{*}{Video} &  &  & \multirow{2}{*}{\checkmark}\\
\multirow{1}{*}{capability} &  &  & \\ \hline

\multirow{1}{*}{3GPP} &  &  & \multirow{2}{*}{\checkmark}\\
\multirow{1}{*}{standardized}   &  &  & \\ \hline
\multirow{1}{*}{Integrated} &  &  & \multirow{2}{*}{\checkmark}\\
\multirow{1}{*}{TN-NTN}   &  &  & \\ \hline
\multirow{1}{*}{VLL} &  & & \multirow{2}{*}{\checkmark}\\
\multirow{1}{*}{operation}   &  &  & \\ \hline
\end{tabular}
}
\label{table:tab6}
\end{table}
Table \ref{table:tab6} compares the proposed 5G ICNS system with the LDACS \cite{ICAO_LDACS_White_Paper}, and legacy VDL Mode 2 systems. 5G ICNS is a promising system that offers better performance and meets future aviation needs. However, further research is required to validate, mature and optimise this innovative solution. In this section, we discuss several future research directions.
\subsection{Evaluation of 5G-Aligned Integrated CNS}
Developing a combination of analytical and simulation models to evaluate the performance of the proposed 5G ICNS  network architecture is a part of our future work. We will apply the realistic communication parameters from 3GPP and ray tracing techniques to accurately model a real-world airport scenario. These analytical and simulation tools will have a modular structure,  each block can be expanded and customised depending on the future needs. We will investigate the proposed 5G ICNS performance for very low-altitude scenarios to deliver the full range of aeronautical communication, navigation, and surveillance services. We will study the QoS of the proposed 5G ICNS by applying the Required Communications Performance (RCP), Required Navigation Performance (RNP), and Required Surveillance Performance (RSP) metrics.
\subsection{Agile Network with Aircraft-Centric Mobility Design}
The envisioned 5G ICNS network aims to connect aircraft to TN infrastructure and NTN, specifically, LEO satellite. This multi-layered integrated TN-NTN architecture requires advanced inter-technology handover mechanisms between different components — for example, the handovers between the aircraft-to-TN segment and the aircraft-to-NTN segment. Specifically, the LEO satellites move rapidly relative to the Earth's surface, limiting their coverage of a given area to brief periods and necessitating frequent handovers in both service and feeder links. This requires a novel solution to efficiently handle handovers. Additionally, the high velocity of LEO satellites induces a significant Doppler effect, which can impair radio channel performance \cite{Asad_Doppler,DtS}. 

\subsection{Network Configuration and Management Policies}
Successful deployment and operation of an ICNS will rely on enabling hybrid networks and interworking between the network segments. As illustrated in Fig. \ref{fig:fig4}, the envisioned 5G ICNS network architecture includes TN and NTN systems, which introduce significant complexity. For example, the integration of TN with NTN would require integration between the cores of the two networks to facilitate the interworking of signalling protocols needed to enable network user devices (aircraft) to efficiently handover from one network segment to the other (TN to NTN and NTN to TN). 
\subsection{Optimization for Aerial Coverage}
To leverage the full potential of the envisioned 5G ICNS, there is a need for further research to optimise the network specifically for VLL aerial corridors, which are defined by air traffic control authorities. The optimisation should consider improving the performance in pre-defined aerial corridors rather than extending ICNS coverage globally. This would require optimising antenna configurations and steering techniques to provide network coverage in air corridors \cite{KarGerJaf2024}. To address the complexity of optimising network performance in 3D space, a promising approach is to leverage data-driven Bayesian optimization to identify configurations that enhance aerial coverage, improve aircraft-cell association, and optimize overall system capacity \cite{BenGerLop2023}.

\subsection{Interference Mitigation and Coexistence} 

The benefits of reusing the same spectral resources for coexisting ground and aerial users may be offset by the mutual interference they generate. 5G-and-beyond cellular networks, which employ massive MIMO beamforming technology, can leverage directionality for spatial separation and interference mitigation \cite{GarGerLop2019}. While air-to-ground communications in mmWave bands are less susceptible to inter-cell interference \cite{KanMezLoz2021}, they consume more power and are highly sensitive to blockage and vibrations. As a result, they may need to operate in a non-standalone mode, such as through sub-6 GHz-plus-mmWave UAS multi-connectivity solutions. Moreover, achieving a sufficiently high link budget at high frequencies will require precise beam alignment and tracking.

\section{Conclusion}
\label{sec:sec6}
In this paper, we discuss the integration of
communication navigation, and surveillance services towards a unified CNS framework that uses integrated 5G TN and NTN systems. We expect that the proposed 5G ICNS could potentially offer a high QoS performance for efficient ATM and UTM collaboration and aviation situational awareness. Our envisioned framework aims to offer robust CNS services while reducing operational costs and improving QoS, resilience and spectrum efficiency. The proposed 5G ICNS framework reduces the needed hardware and network size, which contributes to the minimisation of the wireless systems' CO$_2$ footprint. In addition to safe aviation operation, an efficient CNS system with accurate navigation service plays a crucial role in optimising the flight route, which further reduces fuel consumption, lowers operational costs, and reduces aviation CO$_2$ emissions, which currently account for 2.5\% of global CO$_2$ emissions. Further research is required to evaluate the 5G ICNS performance and optimise the proposed system to ensure that the expected performance for critical services is achieved.  It is also worth investigating the possibilities of A-PNT using the envisioned 5G ICNS system.  

\section*{Acknowledgment}
This work has been performed under the ANTENNAE project, which has received funding from SESAR3 Joint Undertaking under the European Union's HORIZON-SESAR-2023-DES-ER-02 topic, Grant Agreement No. 101167288. However, the views and opinions expressed are those of the authors only and do not necessarily reflect those of the European Union or SESAR 3 JU. Neither the European Union nor the SESAR 3 JU can be held responsible for them.
\bibliographystyle{IEEEtran}
\bibliography{IEEEabrv,ref}
\end{document}